\documentclass[showpacs,twocolumn,prb]{revtex4-1}
\usepackage{graphicx}
\graphicspath{{figures/}} 

\newcommand{\bwt}{\begin{widetext}}
\newcommand{\ewt}{\end{widetext}}

\newcommand{\be}{\begin{equation}}
\newcommand{\ee}{\end{equation}}
\def\bea {\begin{eqnarray}}
\def\eea {\end{eqnarray}}

\usepackage{hyperref}
\hypersetup{
    colorlinks=true,
    linkcolor=blue,
    filecolor=magenta,
    urlcolor=blue,
    citecolor=magenta
}

\usepackage{color} 
\newcommand{\itt}{\it}

\newcommand{\redd}{\textcolor{red}}

\usepackage{comment}
\def\comment#1{}

\usepackage{ulem}   

\begin{document}

\title{Phonon Self-Energy Effects in Migdal-Eliashberg Theory}\thanks{\redd{This paper first appeared in F. Marsiglio, {\itt Phonon Self-energy Effects in Migdal-Eliashberg Theory}, In: {\it Electron--Phonon Interaction in Oxide Superconductors}, edited by R. Baquero (World Scientific, Singapore, 1991) p.167.  Since this source is relatively inaccessible, this
arXiv version has been provided.}}

\author{F. Marsiglio}
\affiliation{Theoretical Physics Branch, Chalk River Laboratories, AECL, Chalk River, Ontario, Canada, K0J 1J0}

\begin{abstract}
Recent work on an electron-phonon model in two-dimensions is reviewed and some new results are presented. We utilize both Quantum Monte Carlo simulations and Migdal-Eliashberg theory. The type of electron-ion coupling considered is on-site. Competing instabilities are charge density wave (CDW) and singlet superconductivity. When the electron band is half-filled, the charge density wave dominates. Away from half-filling CDW correlations actually suppress superconductivity.
\end{abstract}

\date{Dec. 11, 1990}
\maketitle

\section{introduction}

The discovery of high temperature superconductivity has motivated renewed interest in the question of competing instabilities and their relation to superconductivity. In particular, in most of the cuprates the phase diagram shows an antiferromagnetic and superconducting ground state in close
proximity to one another. Even earlier, it had been noted that a charge density wave (CDW) and superconducting ground state competed with one another in BaPb$_{1-x}$Bi$_x$O$_3$. Such competition between various non-Fermi liquid-like ground states has been a recurrent theme in A-15, organic and heavy fermion materials.

In this paper we do not wish to address specifically any of the materials mentioned above. Instead, we wish to consider a highly idealized model and investigate in particular the competition between superconductivity and a charge density wave. The model under consideration is the Holstein
(or molecular crystal) model with Hamiltonian:
\begin{eqnarray}
H  & =  & -\sum_{{\langle i j \rangle} \atop \sigma} t_{ij} (c^\dagger_{i\sigma} c_{j \sigma} + c^\dagger_{j\sigma} c_{i \sigma}) -\mu \sum_{i\sigma} n_{i\sigma} \nonumber \\
& + &  \sum_i \left( {p_i^2 \over 2M} + {1 \over 2} Kx_i^2  \right) -\alpha \sum_{i\sigma} x_i (n_{i\sigma} - {1 \over 2}) 
\label{ham_holst}
\end{eqnarray}
where $x_i$ and $p_i$ are scalar position and momentum operators for the ion with mass $M$ and spring constant $K$ ($=M\omega_E^2$) at site $i$. $c^\dagger_{i\sigma}$ is the creation operator  for an electron with spin $\sigma$ at site $i$. Hopping occurs via the first term in the Hamiltonian, with probability amplitude $t_{ij}$. In this paper, this summation will be restricted to $i$ and $j$ nearest neighbours only. The chemical potential, $\mu$,
is chosen to give a set number density of electrons, $n$. The last term couples the electron density $n_{i\sigma} \equiv c^\dagger_{i\sigma} c_{i \sigma}$ with the lattice displacement, $x_i$ at site $i$, with coupling constant $\alpha$.

A variety of techniques has been used to treat the Hamiltonian in Eq.~(\ref{ham_holst}), most notably, in recent years, Quantum Monte Carlo simulation.
Hirsch and Fradkin\cite{hirsch82} investigated the one dimensional case. The two dimensional case has been investigated by Scalettar er al.,\cite{scalettar89} Levine and Su\cite{levine90} and the author.\cite{marsiglio89} Perturbation theory has also been used and will be further discussed in this paper. The idea is to approach any pending phase transition from high temperature and to determine the susceptibilities for either singlet pairing or charge density wave formation. Other approaches are also possible, such as varying the chemical potential (occupation) or, less physically, the
coupling constant, $\alpha$. Here we review the two dimensional work which has been done including some very recent work by Noack et al.\cite{scalettar89} The strategy is to simulate the Hamiltonian (\ref{ham_holst}) on a small square lattice ($4 \times 4$) while at the same time
we solve the equations arising from self-consistent perturbation theory on the same size lattice with the same boundary conditions (in this case, periodic).
By comparing susceptibilities we get a quantitative measure of the accuracy of our perturbative approximation, which can then be evaluated in the thermodynamic limit. In the following we use units such that $\hbar = k_B = 1$.

\section{Theoretical Background}

The Monte Carlo algorithm used is that of Blankenbecler et al.\cite{blankenbecler81} with modifications due to Hirsch\cite{hirsch88} to give
greater stability at lower temperature. We have used the Monte Carlo results primarily as benchmarks with which to guide our perturbative calculations. 

\begin{figure}[tp]
\begin{center}
\includegraphics[width=3.6in]{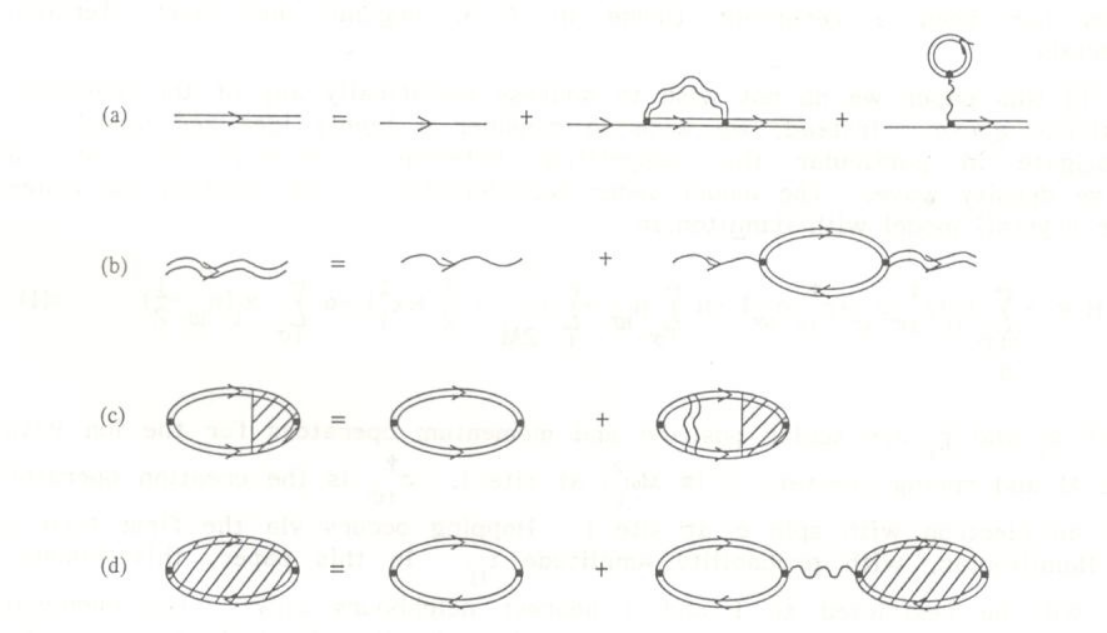}
\end{center}
\caption{Fully self-consistent equations for (a) the interacting electron propagator (straight double lines), (b), the interacting phonon propagator (wavy double line), (c) the singlet pairing susceptibility in the ``ladder'' approximation, and (d) the CDW susceptibility in the ``bubble'' approximation. Single lines
represent non-interacting propagators.}
\label{fig1}
\end{figure}

The perturbative approach is summarized diagrammatically in Fig.~\ref{fig1}. The one-electron Green's function is given by
\be
G(\vec{k},i\omega_m) = [i\omega_m - (\epsilon_k - \mu) - \Sigma(\vec{k},i\omega_m)]^{-1}
\label{e_green}
\ee
where  $i\omega_m \equiv i\pi T (2m-1)$ are the fermion Matsubara frequencies, and $\epsilon_k$ is the single particle energy
obtained from diagonalizing the first term in the Hamiltonian (\ref{ham_holst}). The self-energy due to the electron-phonon interaction
is given by $\Sigma(\vec{k},i\omega_m)$. Similarly the phonon propagator is given by 
\be
D(\vec{q},i\nu_m) = [-M(\omega_E^2 + \nu_n^2) - \Pi(\vec{q},i\nu_n)]^{-1}
\label{p_green}
\ee
where  $i\nu_n \equiv 2i\pi Tn$ are the boson Matsubara frequencies, and $\omega_E$ is the Einstein phonon frequency.
In this model, because we have used a site-diagonal coupling, the unperturbed phonon frequency has no dispersion.
Finally $\Pi(\vec{q},i\nu_n)$ is the phonon self-energy due to the electron-phonon interaction. Migdal\cite{migdal58} originally
argued that $\Pi(\vec{q},i\nu_n)$ led to a renormalization of the phonon frequency $\omega_E \rightarrow \Omega(\vec{q})$. In practice the
phonon spectrum is often taken from experiment, so that the calculation of $\Pi(\vec{q},i\nu_n)$ is not required. Here, since we wish to
predict pending instabilities, it is necessary to retain $\Pi(\vec{q},i\nu_n)$ explicitly.

The approximations indicated in Figs.~\ref{fig1}a,b are:
\begin{eqnarray}
\Sigma(\vec{k},i\omega_m)&  = & {\alpha^2 \over N\beta} \sum_{k^\prime,m^\prime} D(\vec{k} - \vec{k^\prime},i\omega_m - i\omega_{m^\prime}) G(\vec{k^\prime},i\omega_{m^\prime}) \phantom{aaa}
\label{self_e}\\
\Pi(\vec{q},i\nu_n)  & = & {2\alpha^2 \over N\beta} \sum_{k,m} G(\vec{k},i\omega_m) G(\vec{k} + \vec{q},i\omega_m + i\nu_n).
\label{self_p}
\end{eqnarray}
In addition, the occupancy requirement is
\be
n   =  {2\over N\beta} \sum_{k,m} G(\vec{k},i\omega_m) e^{i\omega_m0^+}
\label{occ}
\ee
which determines the chemical potential, $\mu$. Here $\beta \equiv 1/T$. There are no vertex corrections in Eqs.~(\ref{self_e}) and (\ref{self_p}).
However, Eqs.(\ref{e_green}-\ref{self_p}) form a set of {\it fully self-consistent} equations in two dimensional momentum and one dimensional frequency
space, for the electron and phonon propagators. We have demonstrated the need for self-consistently at half-filling previously.\cite{marsiglio89}
The singlet pairing (SP) and CDW susceptibilities are given by:\cite{hirsch85}
\bwt

\begin{eqnarray}
\chi^{\rm SP} & = & {1 \over N} \sum_{ij} \int_0^{\beta} d\tau \langle c_{i\uparrow}(\tau) c_{i\downarrow}(\tau) c^\dagger_{j\downarrow}(0)
c^\dagger_{j\uparrow}(0) \rangle
\label{chi_sp} \\
\chi^{\rm CDW}(\vec{q}) & = & {1 \over N} \sum_{{ij} \atop \sigma \sigma^\prime} e^{i\vec{q} \cdot (\vec{R}_i - \vec{R}_j)} \int_0^{\beta} d\tau 
\left[ \langle n_{i\sigma}(\tau) n_{j\sigma^\prime}(0)\rangle  - \langle n_{i\sigma}(\tau)\rangle \langle n_{j\sigma^\prime}(0)\rangle \right].
\label{chi_cdw}
\end{eqnarray}

\ewt
These susceptibilities are measured directly in the Monte Carlo simulations. In the approximations given in Fig.~\ref{fig1}, they are given by
\be
\chi^{\rm SP}  =  {1 \over N\beta} \sum_{k,m} F(\vec{k},i\omega_m) \Lambda(\vec{k},i\omega_m) 
\label{chi_sp2}
\ee
and
\be
\chi^{\rm CDW}(\vec{q})  = { \bar{\chi}^{\rm CDW}(\vec{q}) \over 1 - \lambda_0 \bar{\chi}^{\rm CDW}(\vec{q}) }
\label{chi_cdw2}
\ee
where
\be
F(\vec{k},i\omega_m) = G(\vec{k},i\omega_m) G(-\vec{k},-i\omega_m),
\label{fdefn}
\ee
\be
\bar{\chi}^{\rm CDW}(\vec{q})  = -{1 \over \alpha^2} {\rm Re} [\Pi(\vec{q},0)]
\label{chi_cdw3}
\ee
and
\be
\lambda_0 \equiv {\alpha^2 \over M\omega_E^2}.
\label{lambda0}
\ee
$ \Lambda(\vec{k},i\omega_m)$ is the solution of the vertex equation:
\bwt

\be
 \Lambda(\vec{k},i\omega_m) = 1- {\alpha^2 \over N\beta}\sum_{k^\prime m^\prime} F(\vec{k}^\prime,i\omega_{m^\prime}) 
 D(\vec{k} - \vec{k}^\prime,i\omega_m - i\omega_{m^\prime}) \Lambda(\vec{k}^\prime,i\omega_{m^\prime}).
 \label{vertex}
 \ee
 
 \ewt
 Equation~(\ref{vertex}) with the ``1'' removed simply becomes the Eliashberg equation to determine $T_c$. Once again iteration
 is required in two dimensional momentum and one dimensional frequency space. 
 
 The importance of including the phonon self-energy, $\Pi(\vec{q},i\nu_n)$ in Eq.~(\ref{p_green}) is apparent upon inspection of Eq.~(\ref{chi_cdw3}).
 In effect we are including the CDW correlations into the one-electron Green's function. In what follows we will call the approximation which includes
 $\Pi(\vec{q},i\nu_n)$ the ``renormalized Migdal-Eliashberg approximation,'' and the one which omits $\Pi(\vec{q},i\nu_n)$ the
 ``unrenormalized Migdal-Eliashberg approximation.''
 
\begin{figure}[tp]
\begin{center}
\includegraphics[width=3.6in]{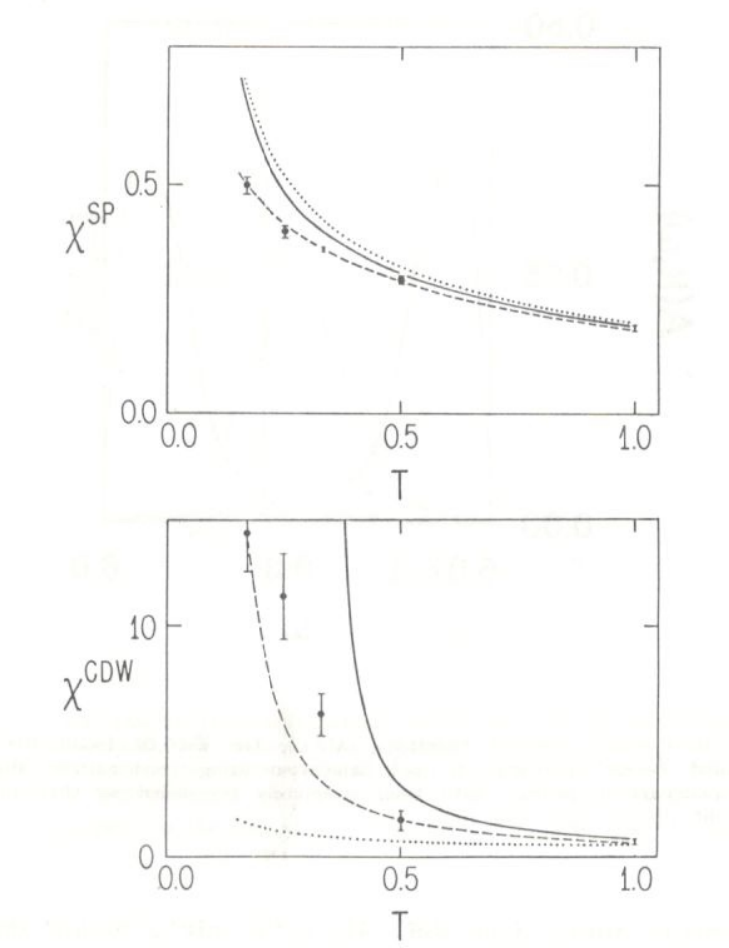}
\end{center}
\caption{Plot of (a) $\chi^{\rm SP}$ and (b) $\chi^{\rm CDW}(\vec{q} = (\pi,\pi))$ vs. temperature for a $4\times 4$ lattice, with $\lambda_0 = 1.5$
and $\omega_E = 0.5$, at half-filling. Points are from Monte Carlo simulations, dashed lines from the renormalized Migdal-Eliashberg approximation and
solid lines from the unrenormalized Migdal-Eliashberg approximation. Including the phonon self-energy self-consistently gives far better
agreement with the Monte Carlo results. The non-interacting limit (dotted lines) is included for comparison.}
\label{fig2}
\end{figure}
\begin{figure}[tp]
\begin{center}
\includegraphics[width=3.4in]{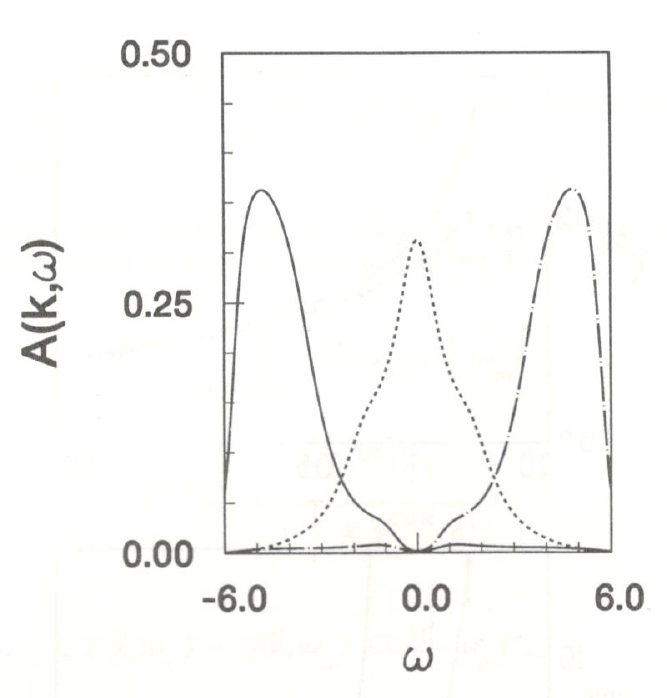}
\end{center}
\caption{Plot of the spectral function $A(\vec{k},\omega)$ for $\vec{k} = (0,0)$ (solid line), $\vec{k} = (\pi/2,\pi/2)$ (dashed line) and
$\vec{k} = (\pi,\pi)$ (dot-dashed line) calculated using renormalized  Migdal-Eliashberg theory. The quasiparticle peaks have been completely broadened
so that the spectra are entirely incoherent ($\lambda_0 = 2$, $\omega_E = 1$, $n=1$).}
\label{fig3}
\end{figure}
\begin{figure}[tp]
\begin{center}
\includegraphics[width=3.4in]{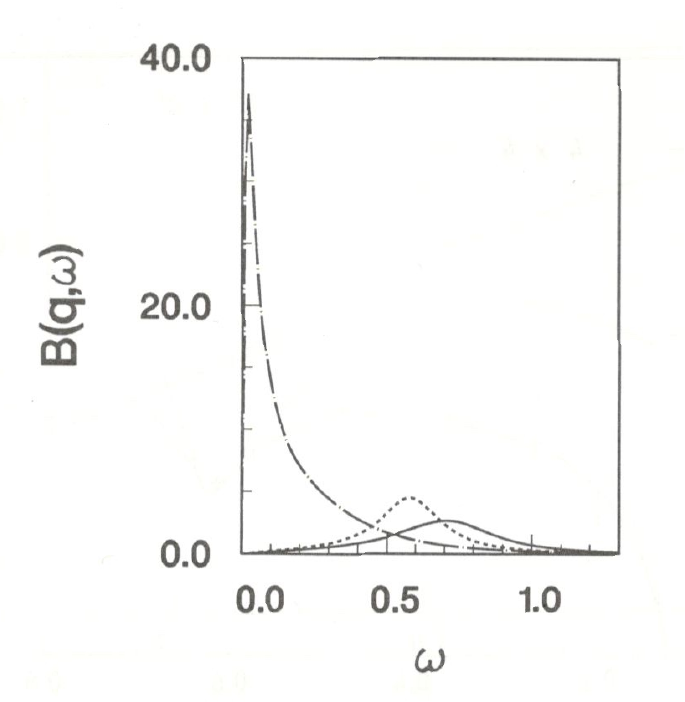}
\end{center}
\caption{Plot of the phonon spectral function $B(\vec{q},\omega)$ for $\vec{q} = (0,0)$ (solid line), $\vec{q} = (\pi/2,\pi/2)$ (dashed line) and
$\vec{q} = (\pi,\pi)$ (dot-dashed line) calculated within the renormalized  Migdal-Eliashberg approximation for the parameters of Fig.~\ref{fig3}.
The unperturbed phonon frequency is $\omega_E = 1.0$. Considerable dispersion has developed, along with mode softening and broadening.
The CDW instability is imminent as the $\vec{q} = (\pi,\pi)$ mode has softened to nearly zero frequency.}
\label{fig4}
\end{figure}
 \section{Half-filling}
 
 The situation at half-filling was investigated in Ref.~[\onlinecite{marsiglio89}], and also recently by Noack et al.\cite{scalettar89} We reproduce in
 Fig.~\ref{fig2} the pairing and CDW susceptibilities as a function of temperature, taken from Ref.~[\onlinecite{marsiglio89}]. Here
 $\lambda_0 = 1.5$ and $\omega_E = 0.5$, in units of the nearest neighbour hopping matrix element, $t$, and we have used a $4\times 4$
 lattice.
It is clear that (a) CDW correlations dominate, and (b) the renormalized Migdal-Eliashberg approximation is required for an accurate quantitative
description. In Fig.~\ref{fig3} we plot the electron spectral function, $A(\vec{k},\omega) \equiv - {\rm Im} G(\vec{k},\omega + i\delta)/\pi$, calculated using
Eqs.~(\ref{e_green}-\ref{occ}) versus frequency for several values of momentum. 
The analytic continuation was determined by using Pad\'e
approximants.\cite{vidberg77} Most noteworthy is the spectral function for $\vec{k} = (\pi/2,\pi/2)$ which is on the Fermi surface, where the spectral function has become entirely smeared out over an energy range of order the original bandwidth. This fact remains for larger lattice sizes as well. This is
to be contrasted with the spectral function computed using $\Pi(\vec{q},i\nu_n) = 0$. In the latter case the quasi-particle peak remains intact (see
Ref.~[\onlinecite{marsiglio89}]), with only a small amount of spectral weight transferred to regions a (unperturbed) phonon frequency away.
This is due to the fact that information of the imminent CDW transition has not been included in the one-electron spectral function in this approximation.
It is thus no longer useful to compute a mass enhancement given by 
$\lambda_k \equiv -\partial {\rm Re} \Sigma(\vec{k},\omega)/\partial \omega|_{\omega = E_F}$. In Fig.~\ref{fig4} we illustrate the phonon spectral function,
$B(\vec{q},\omega) \equiv - {\rm Im} D(\vec{q},\omega + i\delta)/\pi$, for positive frequency for various wave-vectors. Clearly the original
Einstein mode has acquired dispersion with some broadening with the zone corner mode softening nearly to zero frequency. This is another
signal that a $\vec{q} = (\pi,\pi)$ CDW instability is occurring.

\begin{figure}[tp]
\begin{center}
\includegraphics[width=3.5in]{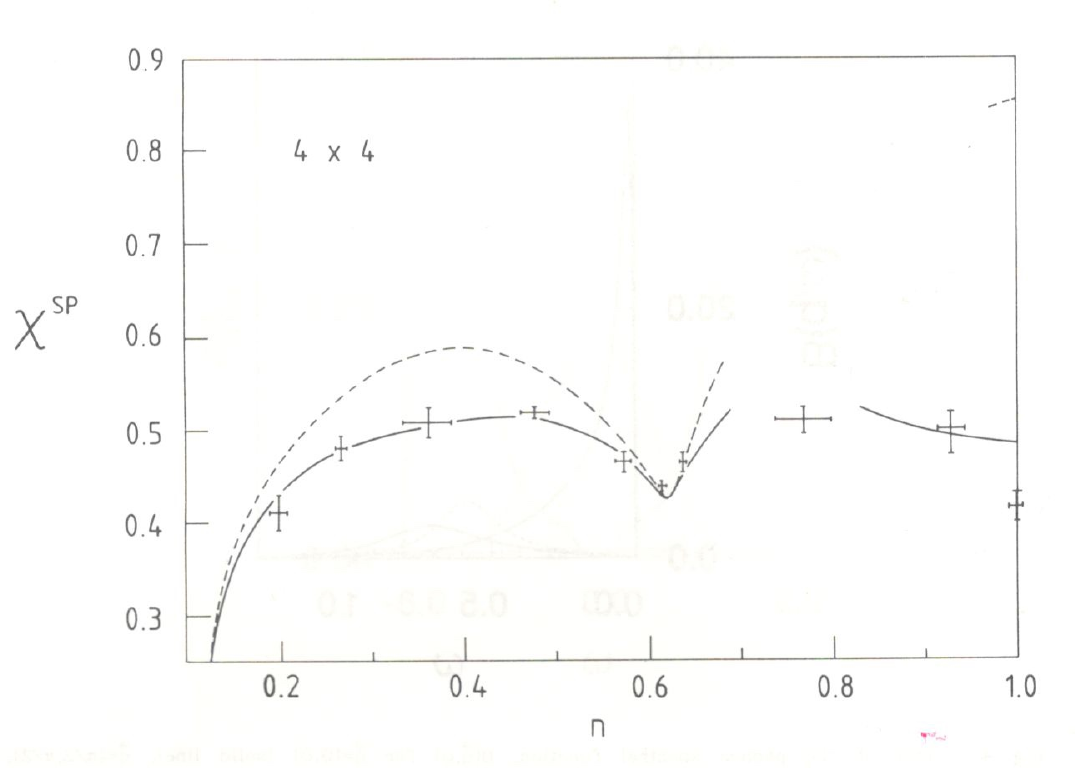}
\end{center}
\caption{Plot of the singlet pairing susceptibility, $\chi^{\rm SP}$ vs. occupation, $n$, for a $4 \times 4$ system 
($\lambda_0 = 2$, $\omega_E = 1$, and $\beta = 6$).  The points with error bars are the Monte Carlo results, the solid line was obtained
using renormalized Migdal-Eliashberg theory  ($\Pi(\vec{q},i\nu_n) \ne 0$) and the dashed line was obtained using  unrenormalized Migdal-Eliashberg theory  ($\Pi(\vec{q},i\nu_n) = 0$). Note the considerable improvement due to phonon renormalization, particularly near half-filling.
Portions of the curves are missing due to the fact that the equations would not converge due to the finite size of the system.
At half-filling the unrenormalized susceptibility vastly overestimates the true result.}
\label{fig5}
\end{figure}

\begin{figure}[tp]
\begin{center}
\includegraphics[width=3.5in]{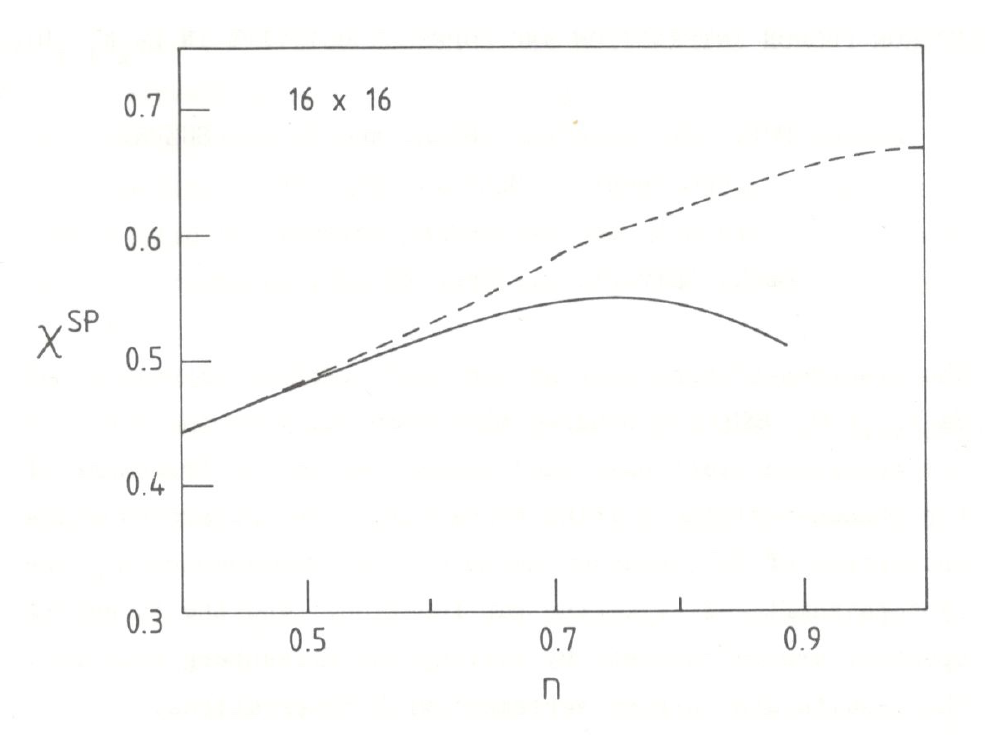}
\end{center}
\caption{Plot of the singlet pairing susceptibility vs. occupation for the same parameters as in Fig.~\ref{fig5} but for a much larger
system size ($16 \times 16$ ), which is representative of the bulk. At low fillings CDW correlations are not present. As the filling approaches 1, the system actually undergoes a transition to a CDW, so that the renormalized Migdal-Eliashberg equations are unable to
converge. (They predict a transition near $n=0.88$ at $\beta = 6$.) It is clear that the strong CDW correlations actually suppress the SP susceptibility at intermediate concentrations.}
\label{fig6}
\end{figure}

\section{Away from half-filling}

At half-filling, the CDW correlations always dominate over the SP correlations. Away from half-filling, the perfect nesting in this model (we have taken
nearest neighbour hopping only) is destroyed and eventually a transition to a superconducting state is expected. The question then 
arises:\cite{levine90,scalettar89} to what extent do CDW correlations enhance or suppress superconductivity? At very low fillings, CDW correlations
are absent, and hence play no role in superconductivity. At intermediate densities, however, they are expected to have some effect. A suitable
test for this hypothesis is to compare our Green's functions calculation with and without CDW correlations with Monte Carlo results.
As already discussed this is achieved by including or omitting $\Pi(\vec{q},i\nu_n)$ in the calculation, respectively. The result for a $4\times 4$ lattice is shown in Fig.~\ref{fig5}. Here, $\lambda_0 = 2.0$, $\omega_E = 1.0$ and $\beta =6$. $\chi^{\rm SP}$ is plotted as a function of density, $n$, obtained
by i) Monte Carlo simulation (points with error bars), ii) the fully self-consistent Green's function  calculation with $\Pi(\vec{q},i\nu_n)$ included (solid
line) , and iii) the Green's function calculation with $\Pi(\vec{q},i\nu_n)$ set equal to zero (dashed line). Finite size effects are quite significant, so all
calculations are done on a $4\times 4$ lattice, with periodic boundary conditions. 
While there is some difficulty in obtaining convergence
near half-filling for the calculations, due to the small system size, it is clear that including CDW correlations in the Green's function calculation provides
far better agreement with the Monte Carlo results.  This gives us confidence that including $\Pi(\vec{q},i\nu_n)$ improves the approximation.
We also take this as a strong indication that CDW correlations influence the pairing susceptibility, and {\it in fact suppress it.} It is difficult
to verify this for much larger lattices, because of the computation time required for the simulations. However, in Fig.~\ref{fig6} we show
the pairing susceptibility for the same parameters as before, but for a $16 \times 16$ lattice, calculated by the two perturbative theories. For $\beta = 6$,
this size lattice is representative of the bulk. Note that in the lower occupation regime in which there previously was a suppression in $\chi^{\rm SP}$
(see Fig.~\ref{fig5}) the two calculations now give essentially the same result. Nonetheless, at occupations closer to half-filling there is clearly a suppression of $\chi^{\rm SP}$ brought on by the CDW correlations. Unfortunately we do not have the Monte Carlo data for a $16 \times 16$ lattice
to verify this. However, based on the success of the renormalized Migdal-Eliashberg result in the $4\times 4$ case (Fig.~\ref{fig5}), Fig.~\ref{fig6}
suggests that CDW correlations suppress the superconducting susceptibility.

\section{summary}

Through a combination of Quantum Monte Carlo simulations on small 2-D systems and fully self-consistent Green's function calculations
using the Migdal-Eliashberg approximation, we conclude that in this model i) at half-filling the ground state is a CDW with ordering
wave-vector $\vec{q}=(\pi,\pi)$, ii) while we have not explicitly demonstrated it here, the ground state at sufficient doping levels away from half-filling will be superconducting, and iii) for concentrations near half-filling the residual CDW correlations actually suppress the SP susceptibility.

\medskip

\begin{acknowledgments}

I would like to thank the authors of Ref.~2b for sending me a preprint of their work before publication. 
The Monte Carlo results were obtained using a program written by J.E. Hirsch.

\end{acknowledgments}

\end{document}